# Coherent force chains in disordered granular materials emerge from a percolation of quasilinear clusters

K.P. Krishnaraj and Prabhu R Nott[*]

*Department of Chemical Engineering, Indian Institute of Science, Bangalore, India*
*Corresponding author: prnott@iisc.ac.in

Abstract

Dense granular materials and other particle aggregates transmit stress in a manner that belies their microstructural disorder. A subset of the particle contact network is strikingly coherent, wherein contacts are aligned nearly linearly and transmit large forces. Important material properties are associated with these force chains, but their origin has remained a puzzle. We classify subnetworks by their linear connectivity, and show the emergence of a percolation transition at a critical linearity at which the network is sparse, coherent, and contains the force chains. The subnetwork at critical linearity closely reflects the macroscopic stress and explains distinctive features of granular mechanics.

Stress transmission in dense, amorphous aggregates of athermal particles, such as granular materials, emulsions, foams and biological cells, is characterized by a spatially inhomogeneous network of interparticle contact forces [1-10]. Numerous experimental [1-10] and computational [11-13] studies have shown that a small subset of the force network, in the form of filamentary quasilinear structures called force chains, transmit large forces. The emergence of a seemingly ordered, spatially correlated network in a structurally disordered particle assembly with only short-ranged repulsive interactions has been a long-standing puzzle. There is substantial evidence that this strong force subnetwork exerts significant influence on the mechanical and transport properties of the medium [10,12,14-18] – understanding its origin and statistical properties is therefore of considerable value.

Previous studies that have analysed force networks in granular materials have primarily aimed at identifying force chains, typically as clusters in which the normal force in all contacts exceeds a threshold [4,5,11,12]. A somewhat more general classification of connected clusters, of which force chains are a specific type, has also been attempted [19]. These studies point to the difficulty of objective identification of force chains based on pair interactions alone. More importantly, they do not offer an explanation for the origin of coherent force transmission in disordered media, and why the spatial correlation of force in force chains is long ranged.

Attempts to understand spatial correlation in granular force networks [13,20] have used the normal contact force $F_n$ to classify subnetworks in grain assemblies subjected to isotropic compression. They found a percolation transition at a critical force $F_n^c$, with critical scaling in the vicinity of $F_n^c$, but came to contradictory conclusions: while Ref. [13] found the scaling exponents to differ from those of the random percolation universality class, and inferred the presence of long-ranged correlation in forces, the more recent study of Ref. [20] investigated larger systems and found the critical exponents to be almost the same as those of the random percolation universality class, thereby inferring the absence of long-ranged correlation. Moreover, these studies do not explain the linear structure of the force chains; indeed, we show in this paper that the structure of the network at $F_n^c$ bears little similarity with experimentally observed force chains.

In this study we use a network connectivity measure to classify subnetworks of connected contacts. Motivated by experimental observations [1-10] that force chains are roughly linearly aligned, we use a simple but robust definition of linearity as the connectivity measure. In computationally generated static granular assemblies subjected to a variety of external forcing, and in the dynamically forced system of steady shear, we find a percolation transition at a critical linearity at which the largest clusters span the system. We show that the clusters of critical linearity constitute the strong force network, of which force chains are a subset, and for which there is long-ranged spatial correlation of the contact force. The orientation of the clusters of critical linearity strongly reflects the imposed macroscopic stress, and explains distinctive, even anomalous, features of the stress in granular columns [21,22]. Finally, we show that linearity percolation is a generic feature of random geometric graphs, thereby explaining the prevalence of force chains in a variety of amorphous particulate systems.

The smallest connected network of particles is a triplet, the linearity of which is $r_t = \boldsymbol{n}_1 \cdot \boldsymbol{n}_2$, where $\boldsymbol{n}_1$ and $\boldsymbol{n}_2$ are the unit normals at the two contacts. We define the linearity of a contact network as

$$r = \min(r_t \mid r_t > 0), \qquad (1)$$

i.e., the minimum triplet linearity in the network such that the angle between adjacent normals is less than π/2 (Fig. 1). It is intuitively apparent that this is a 'weakest link' measure, as the fraction of the normal force from one contact that can be transmitted to the next decreases as $r_t$ approaches 0. With this definition of connectivity, the network of particles in contact is transformed to a weighted graph, with the contacts being the nodes and their connections (triplets) being the edges (see Sec. 2 of Supplemental Material [43]). While this a purely configurational measure of connectivity, we show below that it captures the key features of the strong force network, sheds light on how it arises, and reveals its statistical features.

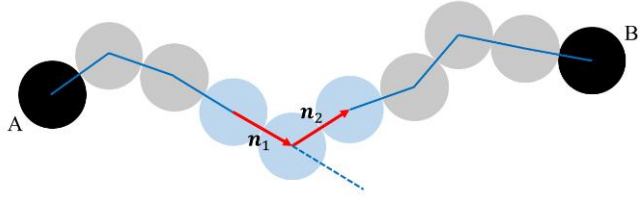

FIG. 1. Definition of cluster linearity. The linearity of the blue triplet is $r_\text{t} = \boldsymbol{n}_1 \cdot \boldsymbol{n}_2$. The linearity $r$ of the cluster is the minimum of $r_\text{t}$ over all triplets in a connected cluster (equation 1). The end particles A and B are either at the boundaries, or in triplets of $r_\text{t} \leq 0$.

We generate configurations of a collection of spheres of mean diameter $d_\text{p}$ from computations using the discrete element method, which computes the motion of the particles using an elastoplastic interaction force. Slight polydispersity in size is maintained to avoid crystalline order. A variety of problems, corresponding to different boundary and forcing conditions were simulated in two and three dimensions. In each problem, the subnetwork of connected triplets of linearity $r$ is sieved out of the entire network, and its connectivity characterized by enumerating the number clusters with $s$ contacts $n(s,r)$ using tools from graph analysis. Details of the computational method, generation of configurations, subnetwork sampling and the determination of network statistics are given in Supplemental Material (SM) [43]. All the results reported are averages over a large number of configurations.

To demonstrate linearity percolation, we first consider a system of $N$ particles in a two-dimensional square domain of size $L$ subjected to isotropic compression. When $r$ is near unity, there are only isolated triplets; as $r$ is decreased, the fraction of connected triplets remains small until a critical linearity $r_c$, at which connected triplets percolate through the system (Fig. 2(a)). The transition exhibits the scaling properties of bond percolation and continuous phase transitions [23]: in the vicinity of $r_c$ the percolation probability $P(r)$ varies as $f(r - r_\text{c})N^{\frac{1}{d\nu}}$, and the mean cluster size $S(r)$ varies as $N^\varphi g(r - r_\text{c})N^{\frac{1}{d\nu}}$, where $\varphi$, $\nu$ are the critical exponents, and $d$ is the dimension of the system (Fig. 2(b)). The values of $r_c$ and the exponents are obtained using standard techniques (see Sec. 2.2 of SM [43]): our estimates are $\varphi = 0.93 \pm 0.29$ and $\nu = 1.15 \pm 0.35$. The large uncertainty in the estimates precludes a conclusion on whether or not it belongs to the random percolation university class [23] ($\varphi = 43/48$, $\nu = 4/3$), but we show below that this has no bearing on the nature of force correlation. It is noteworthy that for the same system, the linearity of force percolated clusters discussed earlier with reference to Refs [13,20] is less than 0.1 (see Fig. S2.2 of SM [43]) – in other words, the clusters at $F_\text{n}^c$ are virtually random, and bear little resemblance to force chains.

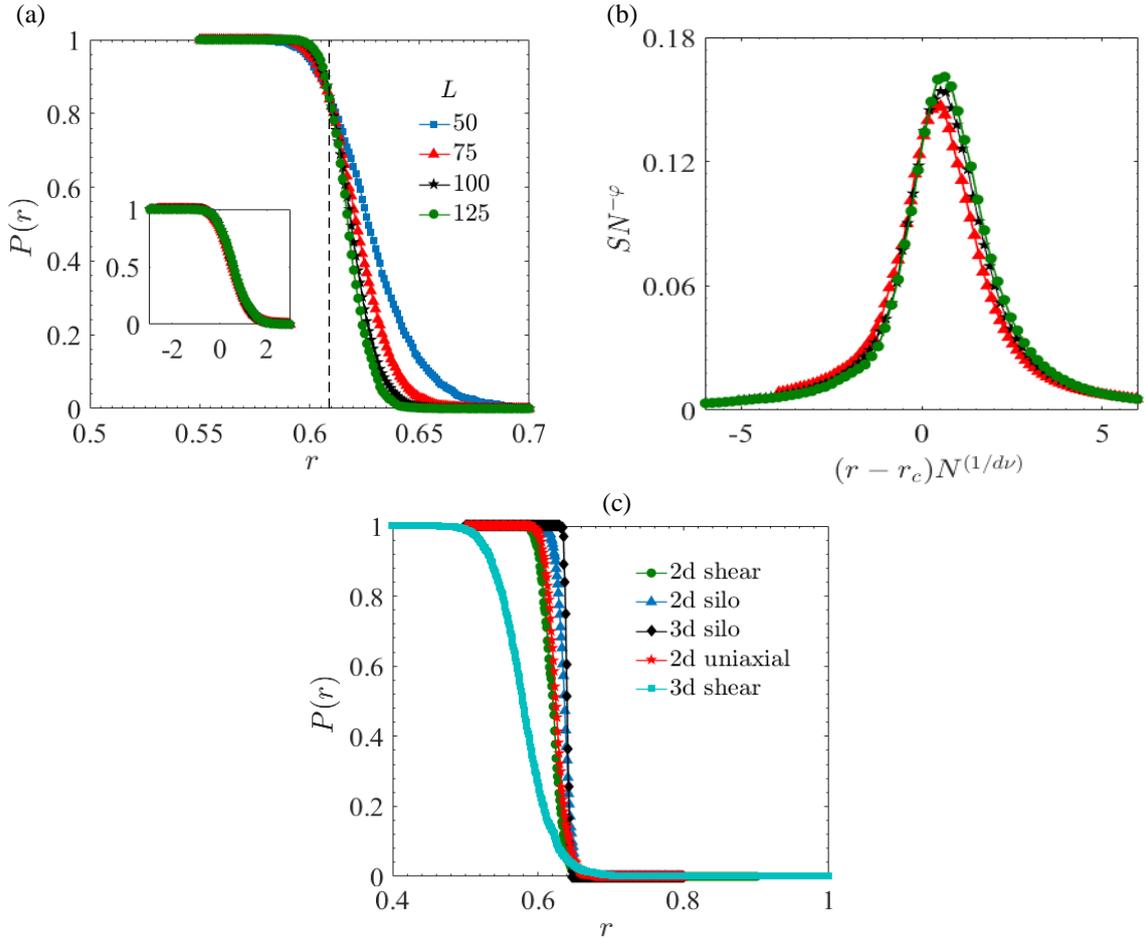

FIG. 2. Linearity percolation. (a) Percolation probability $P(r)$ for 2d isotropic compression for different system sizes $L$ (in units of $d_\text{p}$), the inset showing the collapse for different system sizes when the abscissa is plotted as $(r - r_\text{c}) N^{\frac{1}{d\nu}}$. The dashed line marks the value of $r_\text{c}$, which here is 0.609; it is in general a function of the area fraction. (b) Finite size scaling of the mean cluster size $S(r)$. (c) Percolation probability $P(r)$ for anisotropic forcing, in 2 and 3 dimensions. The geometries and boundary conditions are described in Sec. 3 of SM [43].

Importantly, we observe linearity percolation even for anisotropic forcing, such as static grain assemblies subjected to uniaxial compression and gravity-bound silos, and in the dynamically forced system of steady shear (Fig. 2(c)). This shows that such a connectivity transition is a general signature of granular microstructure in static and slowly deforming states. Estimating the critical exponents for anisotropic systems is not straightforward, and not well studied. However, for our analysis it suffices to obtain an estimate of $r_c$ from the maximum of $\bar{F}_n(r)$ (see Fig. S2.1 of SM [43]); as shown in Fig. 3(b), this provides a good estimate for isotropic compression. We show presently that this estimate of $r_c$ usefully connects the network structure to the macroscopic stress.

Though $r$ is a purely configurational quantity and its percolation is determined without reference to the forces in the contact network, the spatial correlation of force in the network depends strongly on it. To demonstrate this, we compute the correlation function

$$C_F(l) = \langle \delta(l_{ij} - l) F_{n_i}{}' F_{n_j}{}' \rangle \tag{2}$$

where $F_{n_i}{}'$ is the deviation of the normal force at contact $i$ from the mean over all pairs $\langle F_n \rangle$, and $l_{ij}$ is the distance between contacts $i$ and $j$. The angle brackets in (2) denote averaging over the subnetwork of linearity $r$ over multiple configurations. $C_F(l)$ exhibits a power law decay at $r = r_c$ and an exponential decay above or below $r_c$ (Fig. 3(a)), indicating long-ranged correlation at critical linearity, and short-ranged correlation away from it. The mechanical relevance of linearity becomes clearer when we consider the mean force $\bar{F}_n(r)$ in the subnetwork of linearity $r$ – we see that $\bar{F}_n(r)$ is maximum at $r_c$ (Fig. 3(b)). The peak is clearly discernible, but small because the network is sampled from seed contacts chosen randomly; if the seeds are chosen from the subset of contacts that bear a normal force of at least $\langle F_n \rangle$, the mean force in the network rises much more sharply as $r$ increases to $r_c$, followed by a linear rise above $r_c$. The rise in $\bar{F}_n$ for $r > r_c$ is not of mechanical significance, as the radius of gyration of connected clusters drops sharply above $r_c$, and so does their occurrence (see Fig. S2.4 of SM [43]). Thus, clusters of high linearity do not necessarily bear large forces, but the large force bearing clusters are of linearity $r_c$.

A statistical feature of forces in grain assemblies that has been widely observed [2, 3, 12, 24] is that the probability distribution of normal contact force $P(F_n)$ decays exponentially for large $F_n$. In the context of our analysis, it is useful to determine the contact force distribution $P(F_n, r)$ in subnetworks of linearity $r$. This is best illustrated by considering the incremental force network arising from a point force $F_L$ applied on the surface of a gravity-deposited granular bed. The distributions for different $r$ are shown in Fig. 3(c), where it is clear that the probability of finding a large contact force is much higher in subnetworks of linearity $r_c$: specifically, $P(\frac{1}{2} F_L, r_c)$ is over 100 times larger than $P(\frac{1}{2} F_L, 0)$. The results in Fig. 3 provide clear evidence that force chains are subsets of critically linear clusters.

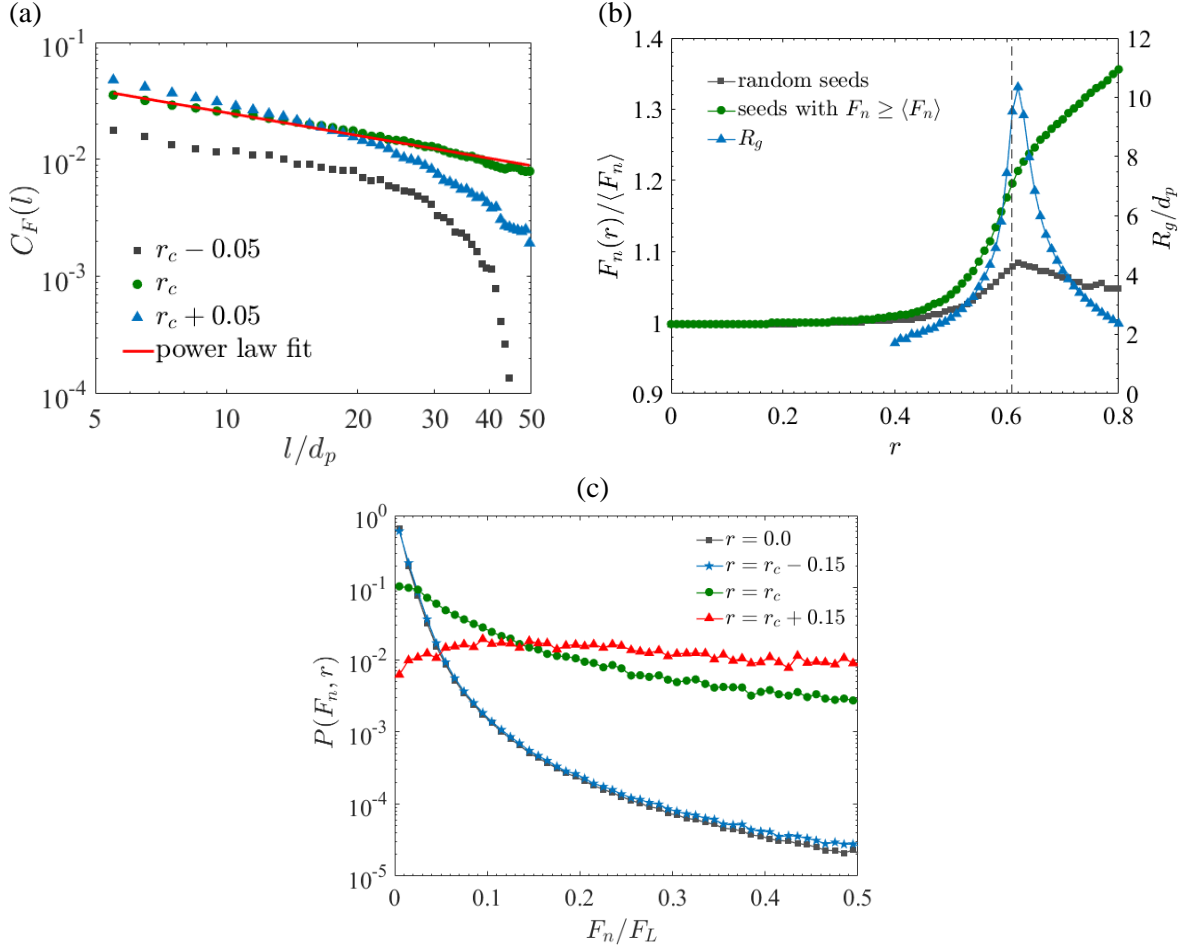

FIG. 3. Statistics of contact forces in subnetworks of different linearity. (a) Spatial correlation function of the contact force $C_F(\ell)$ in the vicinity of critical linearity; the red line is the power law fit $a\ell^b$ at critical linearity ($a = 0.11$, $b = -0.65$). (b) Variation of average contact force with linearity $r$, for random seeds (black squares) and seeds with a minimum contact force of the global mean $\langle F_n \rangle$ (green circles). The radius of gyration $R_g$ of non-percolating clusters is also shown. The dashed line marks critical linearity $r_c = 0.609$. The results in **a** and **b** are for 2d isotropic compression with system size $L = 100 d_p$. (c) Probability distribution of the incremental normal contact force due to a point force $F_L$ acting on the surface of a 2d rectangular bed under gravity (see Sec. 3.4 in SM [43]), in clusters of different linearity; here $r_c$ is 0.66.

Experiments and simulations have shown that force chains in 2d assemblies tend to align along the direction of major principal stress [3, 12, 25]. To probe this feature, we consider the probability distribution $P_{cl}(\theta, r)$ of the orientation of clusters in networks of different linearity (see Sec. 2.4 of SM [43]). For isotropic compression, $P_{cl}(\theta, r)$ too is isotropic, irrespective of $r$ (Fig. 4(a)). In uniaxial compression and plane shear, $P_{cl}(\theta, r_c)$ has sharp peaks in the directions of the major principal stress (Fig. 4(b,c)), but the distribution is nearly isotropic for $r$ even slightly less than $r_c$ – thus, only the subnetwork at $r_c$ closely reflects the anisotropy of forcing. Most previous studies have quantified microstructural anisotropy by examining the distribution of the pair contact vector $\boldsymbol{n}$ (Fig. 1), but it is clear from Fig. 4(b,c) that the orientation of critical clusters is a much more accurate indicator of the macroscopic forcing.

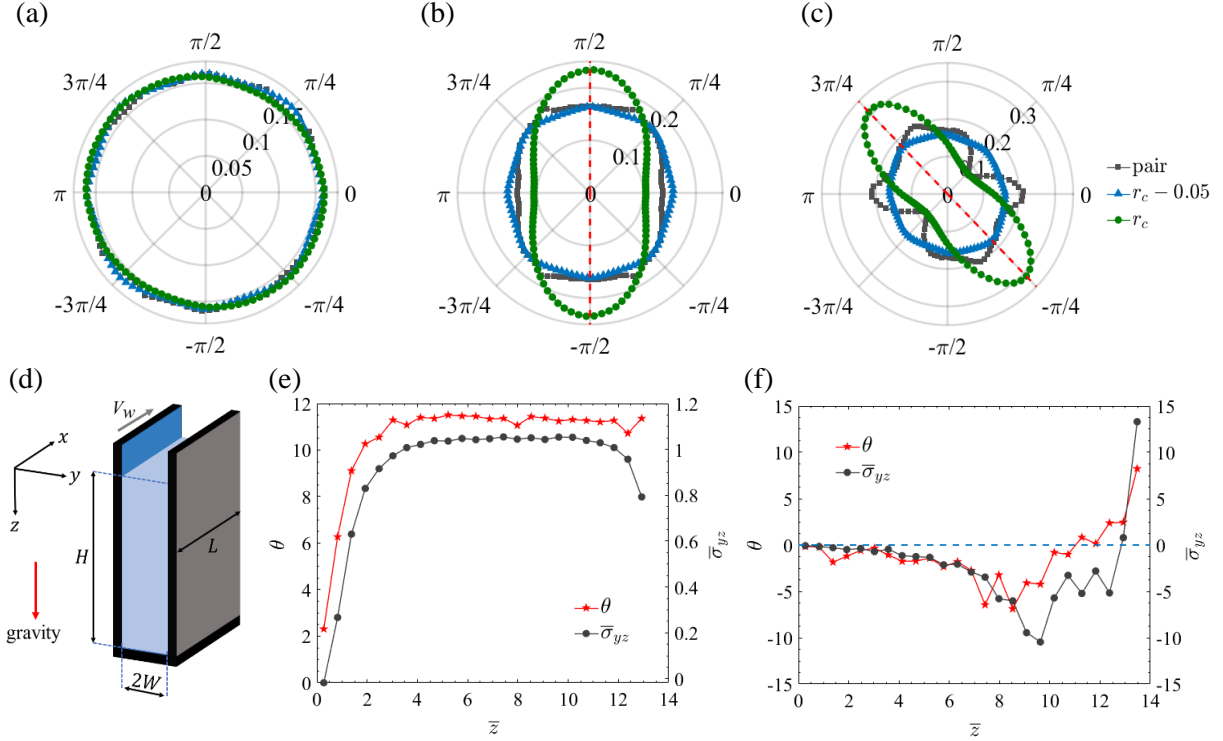

FIG. 4. Clusters of critical linearity reflect the macroscopic stress. (a) Orientation distributions of clusters of different linearity, $P_{cl}(\theta, r)$, and the pair contact vector for different forcing: (a) Isotropic compression ($r_c = 0.6089$), (b) Uniaxial compression ($r_c = 0.63$), and (c) Plane shear ($r_c = 0.62$). In (b) and (c), the red dashed lines indicate the principal directions of compression. (d) Schematic of a 3d granular column; the column is static when the velocity $V_w$ is zero, and it is sheared in the horizontal direction when $V_w$ is non-zero. (e), (f) Variation of the mean critical cluster angle (see Fig. S2.3(b)) and axial shear stress with depth $z$ in a 3d granular column of lateral dimension $2W = 20d_p$ when the column is respectively static, and sheared in the $y$ direction. Here, $\bar{z} \equiv z/(2W)$ and $\bar{\sigma}_{yz} \equiv \sigma_{yz}/(\rho g W)$ are the scaled depth and vertical shear stress.

Indeed, the orientation of critical clusters explains some distinctive and non-trivial features of the stress in granular columns under gravity. It is well known that the stress in a static column of grains saturates exponentially with depth, as a result of a vertical shear stress $\sigma_{yz}$ imposed by the walls owing to Coulomb friction [21, 26]. Recently, a curious feature was revealed when the column is sheared in the horizontal direction [22] by moving one wall relative to the other with velocity $V_w$ (Fig. 4(d)) – $\sigma_{yz}$ changes sign due to a dilation-driven secondary flow [27], and the magnitudes of all stress components rise exponentially with depth. In both these cases, we trace out clusters of critical linearity starting from seed contacts on the walls, and examine their average orientation $\theta$ (see Sec. 2.4 of SM [43]). In a static column, $\theta$ is positive (Fig. 4(e)), indicating that the clusters transmit a downward traction to the walls, or $\sigma_{yz} > 0$. In a sheared column, $\theta$ is negative (Fig. 4(f)), indicating that the critical clusters transmit an upward traction to the walls, whence $\sigma_{yz} < 0$. In both cases, the variation of $\sigma_{yz}$ with depth is closely reflected by that of $\theta$. Thus the intriguing observation of Ref. [27], which is yet to find a complete continuum mechanical explanation, is reflected in the orientation of the clusters of critical linearity. Furthermore, in the sheared column the static base influences the stress at sufficiently large depths ($\bar{z} > 10$), causing $\sigma_{yz}$ to change sign; this too is reflected in the profile of $\theta$ (Fig. 4(f)).

It is pertinent to ask what aspect of grain interactions leads to percolation of linearity? To answer this question, we studied the connectivity of random geometric graphs [28, 29] (see Sec. 4 of SM [43]), wherein edges (contacts) are connected randomly, but keeping the average number of connections per node (particle) fixed. Interestingly, we find linearity percolation in such a graph, with critical exponents for the percolation transition $\varphi = 1.42 \pm 0.037$, $\nu = 0.98 \pm 0.034$ that differ considerably from those of the random universality class. This suggests strongly that linearity percolation, and thereby force chains, arise from topological constraints of the contact network, rather than the details of the interaction force (such as friction) and the balance of force and torque on each particle, as suggested by some studies [4, 5, 13]. It explains why similar force networks are observed in a variety of other aggregates of athermal particles, such as emulsions, foams and living cells [1-9].

In conclusion, we have shown that coherent transmission of force via force chains in disordered granular materials arises from a percolation of quasilinear clusters. The subnetwork at critical linearity, corresponding to the percolation transition, exhibits most of the mechanical and statistical features commonly associated with dense granular materials [3, 12, 21, 22, 25], thereby elucidating the importance of force chains in granular mechanics. Our results throw light on why force chains are seen in many disparate physical systems [1-9]. Our study makes two important connections to current studies on dense particulate materials. The first relates to continuum models, where the need for introducing a fabric tensor in the constitutive relation for the stress [22, 30, 31] has been increasingly felt. While all previous studies have used the pair contact vector to derive a fabric, we make a compelling case for using a fabric based on the orientation of clusters of critical linearity. The second connection relates to the statistics of particle configurations: a long-standing proposal [32] is that all configurations for a fixed volume (that satisfy the constraints of force balance on particles) are equally probable, which has found some computational validation [33,34]. Our results indicate that even if this proposal is valid, only a small subset of the particle assembly that correspond to linearity percolating clusters are of mechanical relevance; studying the statistics of such configurations and the forces therein is therefore significantly more useful.

This work was funded by the Science and Engineering Research Board under grant no. EMR/2016/002817. KPK acknowledges funding from the Ministry of Human Resources Development, India. PRN benefited from discussions with several people during a visit to the Kavli Institute for Theoretical Physics, which was supported in part by the National Science Foundation under grant no. NSF PHY17-48958.

# Supplemental Material

## 1. Particle dynamics simulations

The Discrete Element Method (DEM) is a particle dynamics simulator with an elastoplastic interaction force, which is widely used for computational simulation of granular statics and flow [35]. Our simulations were conducted using the open source molecular dynamics package LAMMPS [36], and the contact model and its DEM implementation are described in Ref. [37]. In DEM the particles are treated as deformable, and their interaction forces are calculated from the normal overlap and tangential displacement post contact. The dissipative interaction is modelled by spring-dashpot modules for the normal and tangential directions (Fig. S1), and an additional Coulomb slider in the latter to incorporate a rate-independent frictional force, an important feature of granular materials. For a pair of spheres $i, j$ of radii $R_i, R_j$ at positions $\boldsymbol{x}_i$, $\boldsymbol{x}_j$ in contact, the overlap is

$$\delta \equiv R_i + R_j - |\boldsymbol{x}_{ij}| \tag{S1}$$

where $\boldsymbol{x}_{ij} \equiv \boldsymbol{x}_i - \boldsymbol{x}_j$; the particles are in contact only when the overlap is positive. The components of the relative velocity normal and tangential to the point of contact are

$$\mathbf{v}_{n_{ij}} = (\mathbf{v}_{ij} \cdot \boldsymbol{n}_{ij})\,\boldsymbol{n}_{ij} \tag{S2}$$

$$\mathbf{v}_{t_{ij}} = \mathbf{v}_{ij} - \mathbf{v}_{n_{ij}} - (\boldsymbol{\omega}_i R_i + \boldsymbol{\omega}_j R_j) \times \boldsymbol{n}_{ij} \tag{S3}$$

where $\boldsymbol{n}_{ij} \equiv \boldsymbol{x}_{ij}/|\boldsymbol{x}_{ij}|$ is the unit normal from $j$ to $i$, $\mathbf{v}_{ij} \equiv \mathbf{v}_i - \mathbf{v}_j$, and $\boldsymbol{\omega}_i, \boldsymbol{\omega}_j$ are the rotational velocities of particles $i$ and $j$. The tangential spring displacement $\boldsymbol{u}_{t_{ij}}$ is initiated at the time of contact and can be calculated by integrating,

$$\frac{d\boldsymbol{u}_{t_{ij}}}{dt} = \mathbf{v}_{t_{ij}} - \frac{(\boldsymbol{u}_{t_{ij}} \cdot \mathbf{v}_{ij}) x_{ij}}{|r_{ij}^2|} \tag{S4}$$

The second term represents rigid body rotation around the point of contact and ensures that $\boldsymbol{u}_{t_{ij}}$ lies in the tangent plane of contact.

For simplicity, the springs are assumed to be linear (Hookean). Previous studies [37] have shown that employing non-linear springs that corresponds to Hertzian contact makes no qualitative difference. The normal and tangential forces imparted on $i$ by $j$ are

$$\boldsymbol{F}_{n_{ij}} = k_n\,\delta_{ij}\,\boldsymbol{n}_{ij} - \gamma_n\,m_{\text{eff}}\,\mathbf{v}_{n_{ij}} \tag{S5}$$

$$\boldsymbol{F}_{t_{ij}} = \begin{cases} -k_t\,\boldsymbol{u}_{t_{ij}} - \gamma_t\,m_{\text{eff}}\,\mathbf{v}_{t_{ij}} & \text{if } |\boldsymbol{F}_{t_{ij}}| < \mu\,|\boldsymbol{F}_{n_{ij}}| \\ -\mu\,|\boldsymbol{F}_{n_{ij}}|\,\mathbf{v}_{t_{ij}}/|\mathbf{v}_{t_{ij}}| & \text{otherwise} \end{cases} \tag{S6}$$

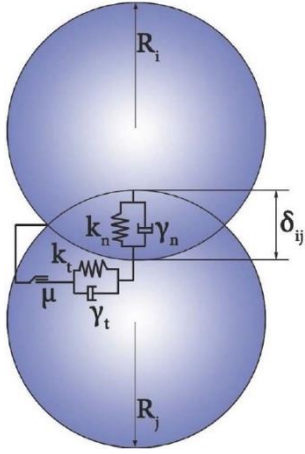

| Parameter | Value |
|---|---|
| $k_n$ | $1.336 \times 10^5 (m_p g/d_p)$ |
| $k_t$ | $\frac{2}{7} k_n$ |
| $\gamma_n$ | $87.82 \, (g/d_p)^{1/2}$ |
| $\gamma_t$ | $\frac{1}{2}\gamma_n$ |

FIG. S1. Schematic of the soft particle interaction model between spheres of radii $R_i$ and $R_j$. The values of parameters used in the model are given in the table, where $g$ is the gravitational acceleration on earth, and $m_p$ is the mass of a particle of diameter $d_p$.

where $k_n$ and $k_t$ are the normal and tangential spring stiffness coefficients, $\gamma_n$ and $\gamma_t$ the corresponding damping coefficients, $\mu$ is the coefficient of friction for the Coulomb slider, and $m_{eff} \equiv m_i m_j/(m_i + m_j)$ is the effective mass of the two spheres. The velocities and positions of the particles are updated by integrating Newton's second law,

$$m_i \dot{\mathbf{v}}_i = \sum_j \mathbf{F}_{ij} + \mathbf{F}_i^{ext}, \quad I_i \dot{\boldsymbol{\omega}}_i = -\frac{\sum_j x_{ij} \times F_{ij}}{2} \quad (S7)$$

where pairwise additivity of the interaction forces is assumed, and $\mathbf{F}_i^{ext}$ is the external force (such as gravity).

For the linear spring-dashpot-slider model, the time of contact is [37]

$$t_{coll} = \pi(2k_n/m - \gamma_n^2/4)^{-1/2}. \quad (S8)$$

The choice of the normal spring stiffness coefficient determines the collision time between two particles. The simulation time step is chosen such that each collision is resolved accurately, and the choice of $\Delta t = t_{coll}/50$ is found to be sufficiently small [27,37]. Since the collision time decreases with increasing spring stiffness $k_n$, it is standard practice to optimize the value of $k_n$ such that it is large enough for the macroscopic behaviour to mimic that of hard particles, and the time step is large enough for the computations to be tractable. The parameters used in the simulations are listed in Fig. S1.

The values of $k_n$, $k_t$ and $\gamma_t$ was chosen based on previous studies [27,37] that have attempted to model hard grains such as glass beads and sand. The value of $\gamma_n$ chosen is such that the normal coefficient of restitution is 0.7. In all our computations, $\mu$ is set to 0.5. The 2d simulations were conducted by placing spheres in a plane and allowing movement only within the plane.

The particle sizes were chosen from a uniform distribution with lower and upper limits of $0.8d_p$ and $1.2d_\mathrm{p}$ respectively, where $d_\mathrm{p}$ is the mean diameter. The walls were constructed with particles of diameter $d_\mathrm{p}$ set in a close packed linear (2d) or triangular (3d) lattice. In all the simulations, the constants characterizing grain-wall interactions are the same as those for grain-grain interactions.

## 2. Subnetwork sampling and characterization

The network of interacting particles is analyzed by considering contacts as the basic units, or nodes. The nodes are connected by edges, which physically correspond to particle triplets (Fig. S2(a-c)). Identification of the nodes and edges transforms the network into a weighted directed graph. It is important that the graph be a directed one so that all permissible nodes are reached from a given seed node. As a result, the triplets that a contact is associated with depends on the direction chosen. For example, the set of triplets that the contact A-B in Fig. S2(a) is associated with depends on whether the contact vector is A→B or B→A. The weight of edge $i$-$j$ is the triplet linearity $r_{ij} \equiv \boldsymbol{n}_i \cdot \boldsymbol{n}_j$, where $\boldsymbol{n}_i$ is unit vector corresponding to node $i$ (Fig. S2(c)).

For a particle configuration with $N_c$ contacts and a given value of the network linearity $r$ (as defined in Eq. 1), the $2N_c \times 2N_c$ adjacency matrix $A$ is constructed [38,39], whose elements $A_{ij}$ represent the connectivity of edges $i$-$j$ (Fig. S2(d)) and are given by,

$$A_{ij} = \begin{cases} 0, & r_{ij} < r \\ 1, & r_{ij} \geq r \end{cases} \tag{S9}$$

An illustration of the procedure applied to a simple representative particle configuration is shown in Fig. S2. It is easy to see that the resulting graph becomes increasingly sparse as the network linearity $r$ increases.

The above-described transformation of the particle packing into a contact based directed graph allows the use of standard tools of graph analysis. For example, the number of contacts that can be reached from a given contact for a fixed value of linearity is estimated by standard search methods used in graphs [38,39]. The commonly used methods to find reachability in graphs are the Breadth First Search (BFS) and Depth First Search (DFS) [38,39]. We have used the BFS method in this study. For a given value of linearity $r$ and a seed contact, the number of edges in the subgraph can be identified. Enumeration of all the subgraphs would require repeating the search for every contact chosen as a seed, which is computationally challenging for large systems. Instead, a large enough fraction of the contacts is randomly chosen as seeds, and the task of enumeration of the subgraphs is reduced to one of sampling. Choosing 10% of all contacts as seeds yields sufficiently accurate statistics. Increasing the percentage of seeds beyond 10% leaves the results unchanged.

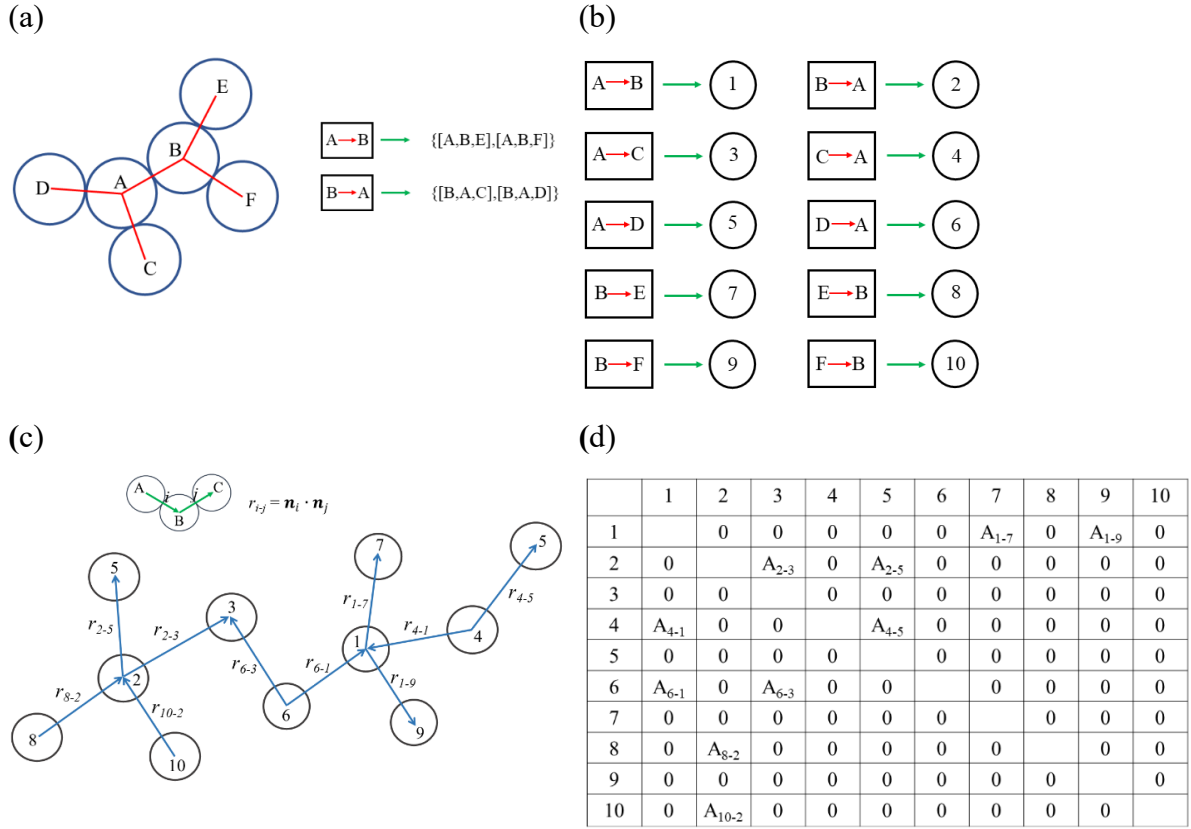

FIG. S2. The cluster identification algorithm. (a) Triplets associated with a contact: the contacts A→B and B→A are connected to different neighbouring contacts in a directed graph. (b) Each directed contact is assigned a unique identity. (c) The weighted directed graph representation of the corresponding particle packing shown in (a). (d) The adjacency matrix representation of (c) in which the elements $A_{ij} = 1$ if the edge $i$-$j$ satisfies the linearity criterion $r_{ij} \geq r$, and 0 otherwise.

## 2.1. Percolation probability and mean cluster size

The percolation probability is found from the subgraphs originating or terminating at wall particles. In isotropic compression, a subgraph is considered percolating if it spans the boundaries of at least a single dimension. In uniaxial compression and plane shear, a subgraph is considered percolating if it spans the non-periodic dimension. In silos, a subgraph is considered percolating if it spans the boundaries in the direction of gravity.

We use the standard definition of cluster number $n(s)$ as the number of subgraphs with $s$ unique contacts. The mean cluster size $S(r)$ then is [23],

$$S(r) = \frac{\sum n_s s^2}{\sum n_s s} \qquad (S10)$$

where the summation is over all $s$. The infinite, or system-spanning, clusters are excluded from the summation.

## 2.2. Estimation of critical linearity and exponents

The value of $r_c$ is obtained from the scaling relation,

$$r_c^{\text{eff}}(L) - r_c \propto L^{1/\nu} \qquad (S11)$$

where $r_c^{\text{eff}}(L)$ is the effective critical linearity of a system of size $L$, and $\nu$ is the correlation length exponent. The estimation of $r_c^{\text{eff}}(L)$ and critical exponents are detailed in Rintoul & Torquato [41] and Pathak et al [20].

The error estimates of the exponents were determined from the 95% confidence interval bounds from regression analysis. The critical exponents and their confidence intervals for isotropic compression were obtained from 2000 configurations, and for random geometric graphs from 50000 configurations.

For anisotropic systems, no clear procedure exists for estimating the critical exponents. For the purpose of this paper it suffices to obtain an estimate of $r_c$, which we get from the maximum of average cluster force $\bar{F}_n(r)$ (Fig. S2.1). Our analysis of isotropic systems (Fig. 3(b)) suggests that this is a good approximation.

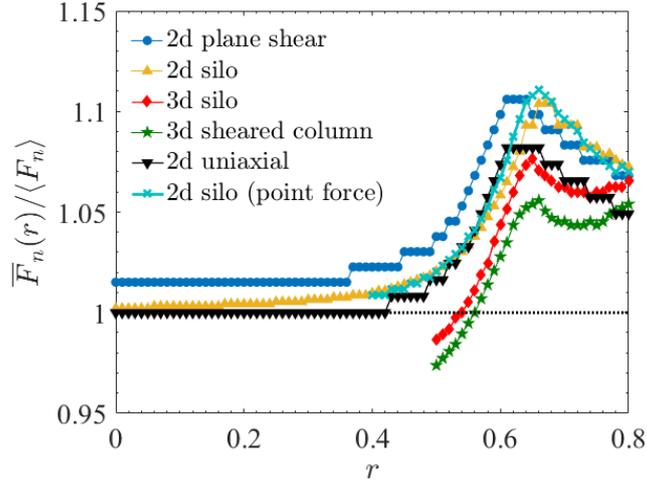

FIG. S2.1. Average cluster force as a function of linearity for anisotropic systems. The linearity $r$ corresponding to the maximum of $\bar{F}_n(r)$ for the different systems are, 0.62 (2d plane shear), 0.67 (2d silo), 0.65 (3d silo), 0.66 (3d sheared column), 0.63 (2d uniaxial compression) and 0.66 (2d silo with point force). The results are averages over 400 configurations for the point force study and 200 configurations for the others.

## 2.3 Linearity of force percolation-based subnetworks

To show the structure of clusters obtained from force percolation, we examined the linearity of system-spanning clusters corresponding to a given threshold force $F_n$. As in Refs [13,20], we isolated the subnetwork of contacts carrying a normal force of $F_n$ or higher, and determined the maximally linear system-spanning path in the subnetwork. To obtain the maximum linearity $r_m$ of a subnetwork, we have used Dijkstra's algorithm [38,39,42] with priority queues. As shown in Fig. S2.2, the maximal linearity of the subnetwork monotonically decreases with increasing $F_n$. We see that the critical force $F_n^c$ at which there is percolation is in the range 1.2–1.5$\langle F_n \rangle$, for which the linearity $r < 0.1$. At such low linearity, the subnetwork is essentially random, and hence the subnetwork at $F_n^c$ bears no resemblance to force chains.

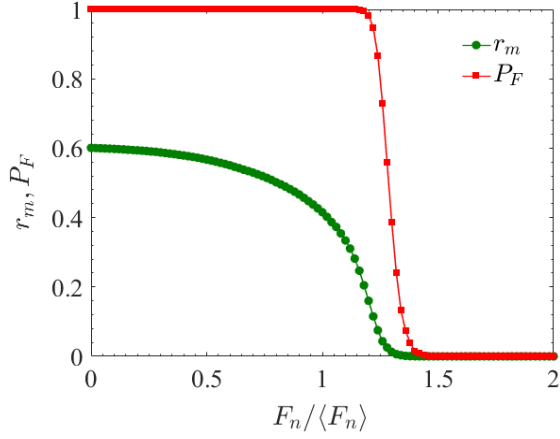

FIG. S2.2. Linearity of force percolation-based subnetworks. Also shown is the percolation probability $P_F$ of clusters whose contacts bear a minimum force of $F_n$. The results are for 2d isotropic compression with dimensions $100 d_p \times 100 d_p$, area fraction of 0.8132, averaged over 2000 independent configurations.

**2.4. Orientation of clusters**

The distribution of the orientation of clusters (Fig. 4(a-c)) is determined in the following manner. For a given value of linearity $r$, a random seed contact is chosen, and the clusters of connected to it are traced. The orientations of the vectors connecting the seed contact to every contact $i$ in the cluster (Fig. S2.3(a)) are then determined; in two dimensions the angle $\theta$ with one coordinate axis determines the orientation. This is repeated over many seeds and multiple configurations, and the probability distribution $P_{cl}(\theta, r)$ of the orientation of such vectors is determined. $P_{cl}(\theta, r)$ is defined unambiguously if the sampling is conducted over every contact chosen as a seed; in practice, we find that randomly choosing 10% of the total number of contacts as seeds in each configuration, and averaging over a sufficiently large number of configurations, provides accurate statistics. The results shown in Fig. 4(a-c) are averages over 200 configurations.

The orientation of clusters in a gravity-bound vertical column is determined in a slightly different manner. Here, the mean orientation of clusters of linearity $r$ connected to the walls in static and sheared vertical columns was determined by choosing contacts with wall particles as the seeds. For each wall particle $j$, the clusters emanating from its contacts are traced, and the angle $\theta_i$ subtended by the line connecting the wall particle with particle $i$ in the clusters with the horizontal (in the clockwise direction) is determined (Fig. S2.3(b)). The average orientation of clusters $\theta$ is

$$\theta = \frac{\sum_{j=1}^{N_\text{w}} \frac{1}{N_\text{c}} \sum_{i=1}^{N_\text{c}} \theta_i}{N_\text{w}} \tag{S12}$$

where $N_\text{c}$ is the number of particles in the critical clusters emanating from the wall particle $j$, and $N_\text{w}$ is the number of wall particles in the vertical strip $\Delta z$ connected to at least one cluster. We average over vertical bins of width $\Delta z \approx 11 d_\text{p}$ and over many configurations to determine the $\theta(z)$ profiles in Fig. 4(e,f). In the sheared column, the critical cluster angles were measured along the stationary wall. The results shown in Fig. 4(e,f) are averages over 200 and 3000 configurations for the static and sheared columns, respectively.

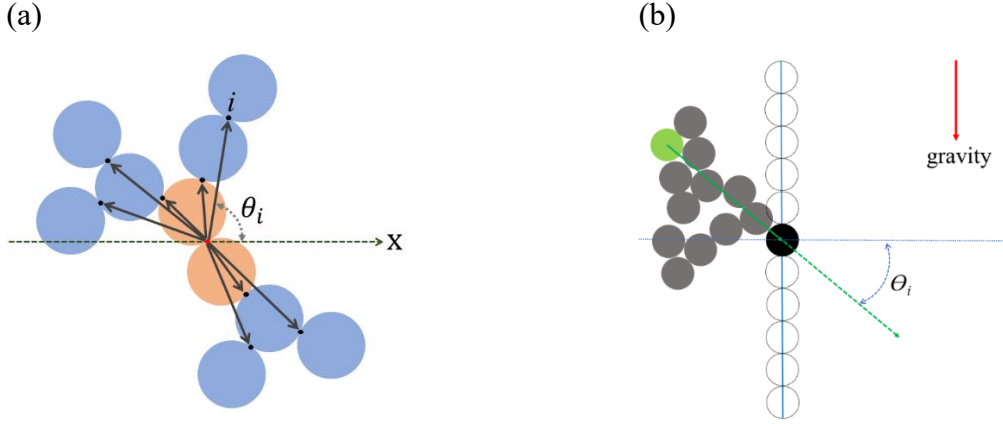

FIG. S2.3. (a) Schematic of cluster orientation estimation. For each value of the linearity $r$, the orientation of the vectors joining a seed contact (light brown particles) to all contacts in the cluster are measured with respect to an arbitrary coordinate axis. The orientations are collected over a sufficiently large number of seeds in each configuration and averaged over many configurations to obtain the probability distribution. (b) Schematic of cluster orientation estimation in a silo. Here the seeds are contacts with wall particles (black) within a vertical strip $\Delta z$, and the angle $\theta_i$ with respect to the horizontal (in the clockwise direction) between the line connecting a particle $i$ in a cluster (green) and the wall particle is measured.

## 2.5. Radius of gyration and size of subnetworks

To study the size of clusters of different linearity, we computed their radius of gyration $R_\text{g}$ as a function of $r$. For a cluster of linearity $r$, the radius of gyration is

$$R_\text{g} = \sqrt{\frac{1}{N_c} \sum_{i=1}^{N_c} (\boldsymbol{x}_i - \boldsymbol{x}_\text{cm})^2} \tag{S13}$$

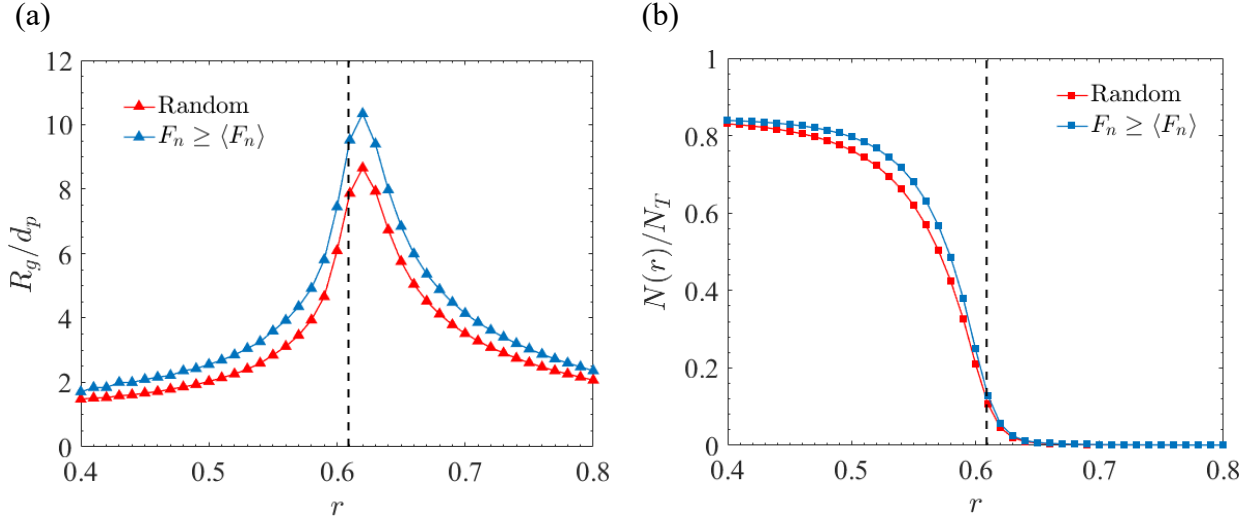

FIG. S2.4. (a) Radius of gyration $R_g$ of non-percolating clusters. (b) Number of particles $N(r)$ in clusters of linearity $r$ scaled by the total number of particles $N_T$. The results are for 2d isotropic compression with dimensions $100 d_p \times 100 d_p$, area fraction of 0.8132, averaged over 200 independent configurations. The dashed line marks the value of $r_c$, which here is 0.609.

where $N_c$ is the number of particles in the cluster, $\boldsymbol{x}_i$ is the position vector of particle $i$ and $\boldsymbol{x}_{cm}$ is the position vector of the centre of mass of the cluster; this is then averaged over all non-percolating clusters of linearity $r$. Figure S2.4(a) shows $R_g$ as a function of linearity $r$ for 2d isotropic compression ($N_T = 12{,}500$, $L = 100 d_p$). The choice of seeds does not have a significant influence on $R_g$, unlike the mean force in the cluster (Fig. 3a). The number of particles $N(r)$ in clusters of linearity $r$ is shown in Fig. S2.4(b), wherein it is clear that the occurrence of clusters of linearity greater than $r_c$, and therefore large $R_g$, is negligibly small.

## 3. Creation of static and sheared configurations
### 3.1. Isotropic compression

To generate the configurations, we start from an initial random configuration of non-overlapping spheres in a 2d square box, whose walls are composed of array of spheres of mean diameter $d_p$. In the initial state, the area fraction is 0.1 and the minimum distance between any two particles is $1.5 d_p$. From the initial state, a force $F$ is applied on the walls, which is increased linearly with time from zero to $F_{\max}$ in a period of $940 \sqrt{\frac{d_p}{g}}$, after which the simulations are continued at force $F_{\max}$ until the kinetic energy per particle in the system decays to $\approx 10^{-11} m_\text{p}gd_\text{p}$. This process is repeated to create multiple configurations. The results for isotropic compression are for an applied stress of $\sigma \equiv F_\text{max}/(Ld_\text{p}) = 2.45\ m_\text{p}g/d_\text{p}^2$ (Fig. 2) and area fraction is 0.8132; decreasing or increasing $\sigma$ causes a corresponding change in the critical linearity $r_\text{c}$ (Fig. S3.1(c)). The results shown in Fig. 2(a,b) are averages over 2000 configurations, and in Fig. 3(a,b) over 200 configurations.

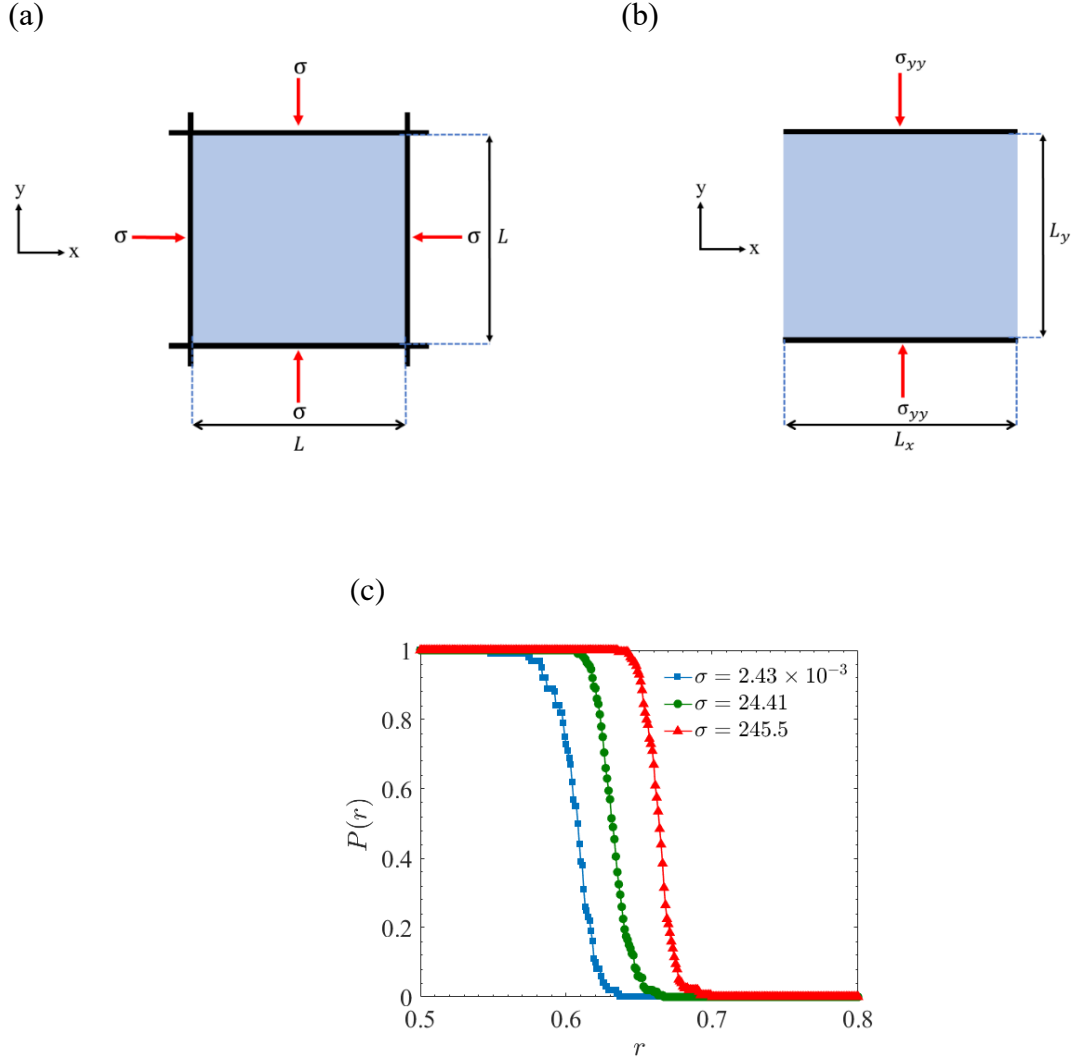

FIG. S3.1. (a) Schematic of 2d isotropic compression. (b) Schematic of 2d uniaxial compression. The left and right boundaries are periodic. (c) Linearity percolation in 2d isotropic compression for increasing boundary stress $\sigma$ (in units of $m_\text{p}g/d_\text{p}^2$). The area fractions of the systems in the order of increasing $\sigma$ are 0.811, 0.817 and 0.827 respectively. The results shown are averages over 200 configurations.

## 3.2. Uniaxial compression

The initial loose packed configuration and the final compressed configurations are obtained in the same manner as in isotropic compression, with the sole difference that the left and right boundaries are periodic. The $x$ dimension of the rectangular box is fixed at $L_x = 100d_p$ (Fig. S3.1(b)), and the $y$ dimension $L_y$, bounded by the rigid walls, is varied to reach the desired initial state. From the initial loose state, a total force $F$ is applied on the two walls, which is increased linearly with time from zero to $F_{max}$ in a period of $940\sqrt{\frac{d_p}{g}}$, after which the simulations are continued with constant force $F_{max}$ until the kinetic energy per particle in the system reaches $\approx 10^{-11} m_p g d_p$. For the uniaxial system shown in Fig. 2(c) and 4(b), $\sigma \equiv F_{max}/(Ld_p) = 0.75\ m_p g/d_p^2$. In the nearly static final state, the $y$ dimension $L_y \approx 100d_p$, and area fraction is 0.811. The results in Fig. 2(c) are averages over 300 configurations for all the geometries and forcing.

## 3.3. Plane shear

In 2d plane shear a monolayer of particles was sheared by restricting their centres from moving out of the $x$-$y$ plane, with periodic boundaries in the $x$ direction (Fig. S3.2(a)). The simulations were started from an initial loose configuration generated in the same manner as in uniaxial compression; thereafter, in addition to applying compressive forces on the two walls, the walls were translated at velocities $V_w$ and $-V_w$ in the $x$ direction. The normal force was increased linearly from zero to $F_{max}$ in a time of $1880\sqrt{\frac{d_p}{g}}$, after which the simulations were continued with constant force and wall velocity for strain $\gamma \approx 30$; the total kinetic energy of the system reaches a steady state value after $\gamma \approx 10$. At steady state, the dimensions of the system are $L_x = 100d_p, L_y \approx 100.5d_p$, area fraction is 0.8133, and normal stress is $\sigma_{yy} \equiv F/(L_x d_p) = 0.75\ m_p g/d_p^2$.

The configurations for 3d shear were created in the same manner, but allowing the particles to translate and rotate in all 3 directions. Periodic boundaries are imposed in the $x$ and $z$ directions (Fig. S3.2(b)). While the 2d shear simulations were performed at constant normal stress $\sigma_{yy}$, the 3d shear simulations were performed at constant volume – the compression of the loose configuration was stopped upon reaching a volume fraction $\phi$ of 0.595, after which the z-coordinates of the walls were fixed, and shearing was commenced by moving the top and bottom walls at constant velocities of $V_w$ and $-V_w$, respectively. The steady state properties were determined after shearing for a strain of $\gamma \approx 100$. The dimensions of the system at steady sheared state are $L = 30d_p$, $D = 20d_p$, $W = 21d_p$.

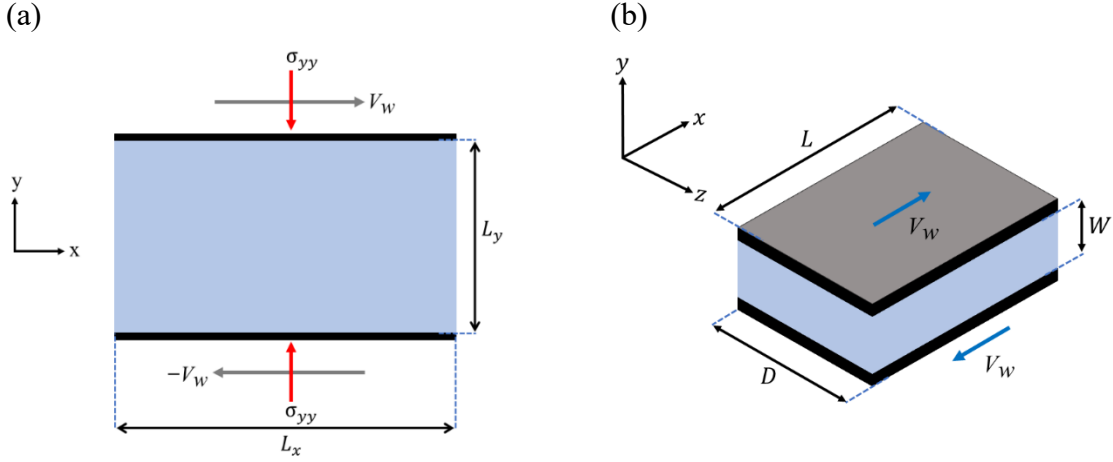

FIG. S3.2. Schematic of plane shear in (a) 2d and (b) 3d.

The regime of flow is characterized by the Savage number [40], defined as the ratio of stress due to grain inertia to the total stress,

$$Sa = \frac{\rho_p d_p^2 \dot\gamma^2}{\sigma_{yy}} \qquad (S14)$$

where $\rho_p$ is the intrinsic density of the particles and $\dot\gamma$ is the shear rate. The inertia number $I$ used in many recent studies is $Sa^{1/2}$. The Savage number for 2d shear flow is $\approx 10^{-8}$, thereby implying that granular material is in the slow flow, or quasistatic, flow regime. The Savage number for 3d shear flow is $\approx 10^{-4}$, at which effects of grain inertia are expected to be small but finite – however, the qualitative features of the contact network are found to be the same as for lower $Sa$.

### 3.4. Silo and point force simulations

For all the studies pertaining to the analysis of linearity percolation and contact force statistics in silos (Figs 2(c) and 4(e,f)), a vertical container of rectangular cross section was filled by 'raining' the grains into it from above under the influence of gravity. The walls were constructed of particles of diameter $d_p$ in exactly the same manner as in the 2d and 3d systems described above. To fill a silo of dimensions $L \times 2W \times H$, particles were created randomly in a pouring region placed above the silo to a volume fraction of 0.01, and allowed to fall into the silo under the influence of gravity. The dimensions of the pouring region were $(L - d_p) \times (2W - d_p) \times H_r$, where $H_r$ was typically between $H/2$ and $H/3$, and its base was 5-$10d_p$ above the free surface of silo (Fig. S3.3). The creation of particles in the pouring region and their raining was continued until the container was filled to the required fill height. To achieve a static state, the simulations were continued until the kinetic energy per particle reduced to $10^{-12}\, m_p g d_p$. This process was repeated to create multiple configurations. The dimensions of the 2d silo are $2W = 40 d_p$, $H = 490 d_p$ (Fig. S3.3(a)), and the particles centres are constrained from moving out of the in the $x$-$y$ plane. The dimensions of the 3d silo are $2W = 20 d_p$, $L = 25 d_p$, $H = 265 d_p$ (Fig. S3.3(c)), with periodic boundaries in the $x$ direction.

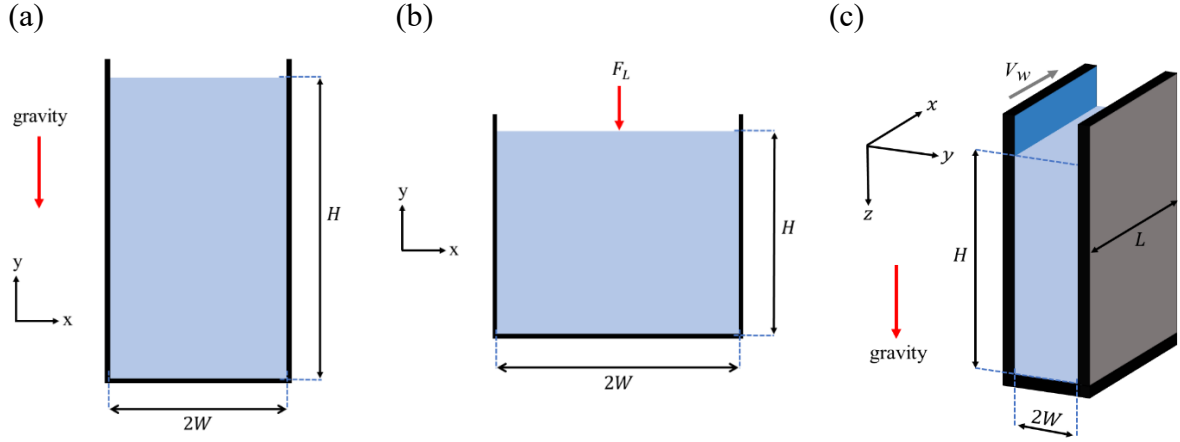

FIG. S3.3. Schematics of the silo and the 2d point force simulations. (a) 2d silo. (b) Point force acting at the symmetry axis on the free surface of the 2d silo. (c) 3d silo; the wall velocity $V_w$ is zero in the static case.

For the point force simulation (Fig. S3.3(b)), a static 2d bed of dimensions $2W = 100 d_p$, $H = 50 d_p$ was first created, and a point force $F_L = m_p g$ was applied to the particle closest to the midpoint on the free surface; the particle positions and forces were further evolved till the kinetic energy per particle reduced to $10^{-12} m_p g d_p$. The response to the point force was measured by determining the change in the force $\Delta F_n$ at each contact upon application of the point force. Due to plastic rearrangements, the normal forces either reduce or vanish in a small fraction of contacts; for our analysis, only the contacts for which for $\Delta F_n > 0$ were considered. The network statistics shown in Fig. 3(c) was obtained from 2000 configurations.

For the analysis of the force network in a sheared column (Fig. 4(f)), static configurations were first created in the same manner as that of a static 3d silo, discussed above. Subsequently the column was sheared, by moving the left wall (Fig. S3.3(c)) with constant velocity $V_w$ in the $x$ direction, for strain $\gamma \approx 50$. The bed dilates during shear, and its steady state dimensions were $2W = 20d_\mathrm{p}, D = 25d_\mathrm{p}, H = 275d_\mathrm{p}$. The vertical shear stress on the wall $\sigma_{yz}$ and cluster angles $\theta$ were then obtained by averaging over the steady state configurations. The shear stress $\sigma_{yz}$ is given by

$$\sigma_{yz} = \frac{\sum_i F_z^i}{L\,\Delta z}, \tag{S15}$$

where $\Delta z$ is the width of the segment of the wall at height $z$, and $F_z^i$ is the vertical component of the force transmitted to the wall particle $i$ in the segment. We chose $\Delta z \approx 11 d_p$ to determine the variation of $\sigma_{yz}$ with depth $z$ (Fig. 4(e,f)). In the sheared column, the axial stress was measured along the stationary wall.

## 4. Random geometric graphs (RGG)

Random geometric graphs are random graphs embedded in Euclidean space [28,29]. RGGs are used to model spatial networks and in studies of continuum percolation [28,29]. We generate random graphs on a 2d square domain of dimension $L$. The positions of $N$ points in the domain are randomly generated, and any two points are considered connected if they are separated by a distance $\leq R$, where $R$ is referred to as the connection distance. The average number of connections per point $\langle k \rangle$ in 2d is given by [28,29],

$$\langle k \rangle = \rho \pi R^2 \tag{S16}$$

where $\rho$ is the density of points in the area considered. The number of points to be generated is found using the values of $\rho$, and area of the domain, $A$. The value of $\rho$ used in this study is 0.6363, and $R$ was chosen such that the average coordination number (number of connections per point) $\langle k \rangle$ is 6. In this graph the points are the nodes and the connections are the edges. The percolation probability and finite size scaling analysis of RGG is shown in Fig. S4. Our estimate of critical linearity for the RGG is $0.646 \pm 0.0009$.

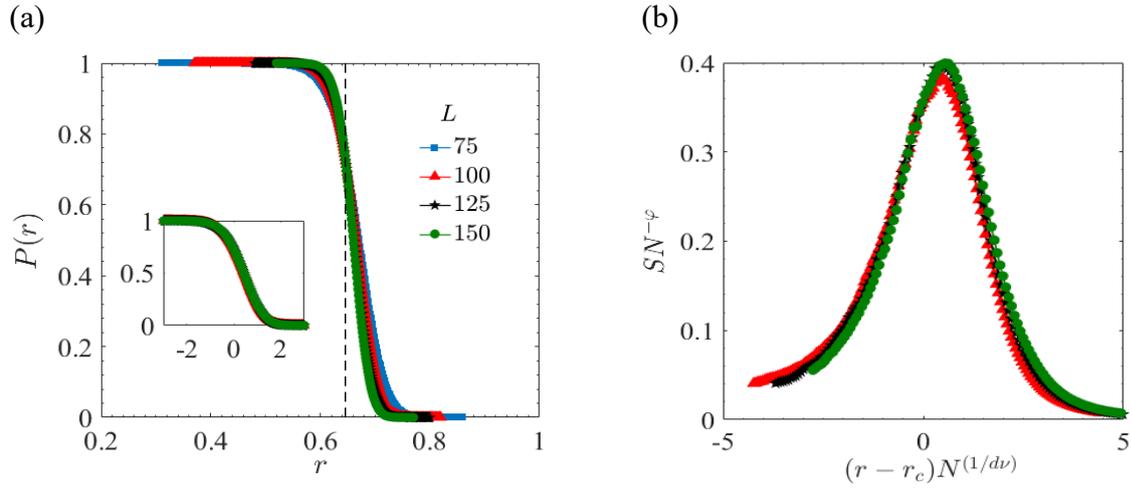

FIG. S4. Finite size scaling in random geometric graphs (RGG). The system size $L$ is in units of the connection distance $R$. (a) Percolation probability for different system sizes; the inset shows collapse of all the data when $P(r)$ is plotted against $(r - r_c) N^{\frac{1}{d\nu}}$ (see Fig. 2). (b) Finite size scaling of the mean cluster size $S(r)$. The results are obtained from 50000 configurations.